\documentclass[%
 reprint,
 amsmath,amssymb,
 aps,pra]{revtex4-1}

\usepackage{color}
\usepackage{graphicx}
\usepackage{dcolumn}
\usepackage{bm}

\newcommand{\C}{\mathbb{C}}

\newcommand{\spc}[1]{\mathcal{#1}}

\def\d{{\rm d}}

\def\>{\rangle}
\def\<{\langle}

\newcommand{\bs}[1]{\boldsymbol{#1}}

\newcommand{\map}[1]{\mathcal{#1}}
\newcommand{\Tr}{\operatorname{Tr}}

\def\qed{$\blacksquare$ \newline}

\begin{document}

\preprint{APS/123-QED}

\title{Quantum amplification and purification  of noisy coherent states}

 \author{Xiaobin Zhao$^{1,2}$ and Giulio Chiribella$^{1,2,3}$}
 \affiliation{$^1$ Department of Computer Science, The University of Hong Kong,
Pokfulam Road, Hong Kong}
\affiliation{$^2$ The University of Hong Kong Shenzhen Institute of Research and Innovation, Kejizhong 2$^{\rm nd}$ Road,  Shenzhen}
\affiliation{$^3$Canadian Institute for Advanced Research, CIFAR Program in Quantum Information Science}

\nopagebreak

\begin{abstract}

Quantum-limited amplifiers increase the amplitude of quantum signals at the price of introducing additional noise. Quantum purification protocols operate in the reverse way, by reducing the noise while attenuating the signal.  Here we investigate a scenario that interpolates between these two extremes. We search for the optimal physical process that generates   $M$ approximate copies of  pure and amplified coherent state, starting from $N$ copies of a noisy coherent state with Gaussian modulation.  We prove that the optimal deterministic processes are always Gaussian, whereas non-Gaussianity powers up  probabilistic advantages in suitable parameter regimes.  The optimal processes  are experimentally feasible,  both in the deterministic and in the probabilistic scenario.   In view of this fact, we provide benchmarks that can be used to certify the experimental demonstration of the quantum-enhanced amplification and purification of coherent states.

\end{abstract}
\maketitle

\section{introduction}

Coping with noise is a fundamental problem in quantum communication networks, where the quality of the communication is  often affected by imperfections in the transmission line, by the presence of eavesdroppers, and by the use of  non-ideal repeaters.   Various techniques have been developed to fight noise: error correction codes focus on  preventing noise \cite{shor-1st-error-corr,QEC-meas,err-corr-2015-rmp}, while purification  techniques can be used to distill cleaner resources from noisy systems \cite{qubit-purification,duan-entpuri,pan-entpuri,fiurasek,puri-exp-coherent-state,puri-prob-coherent-state}.  Purification techniques are crucial to quantum repeaters \cite{QR-QP1,QR-QP2,QR-QP3}, where they can be used to enhance the quality of quantum communication at the price of a reduced rate.  

Another fundamental primitive in quantum optics  is the amplification of quantum signals \cite{caves1982amplification}. Here the task is to increase the amplitude of a weak signal, in order to make it more easily detectable.  This task cannot be achieved perfectly, because a perfect amplification would violate  fundamental quantum principles, such as Heisenberg's uncertainty and the no-signalling principle. The price  for amplification is an increased level of noise, which manifests itself in the form of increased fluctuations of the canonical quadratures.   Still, the price may be worth paying when the original signal is so weak that it would be hard to detect otherwise.  
  The amplification of quantum signals encoded in pure  states has been studied in depth, both theoretically   \cite{ralph-lund-nondeterministic,namiki2011fundamental,caves2012quantumlimits,chiribellaxie,caves2013quantumlimits,caves2014noise,namiki2015amplification,caves2016models} and experimentally \cite{ferreyrol2010implementation,heralded-amp,usuga2010noise,noiseless-amplifier,kocsis2013heralded,bruno2013complete}.      Comparatively little is known, however, about the case of states that have been subject to noise \emph{before} the amplification process.   The goal of this paper is to identify the quantum processes that achieve the best amplification performance and to give criteria for assessing their experimental demonstration.

We will focus our attention on the amplification of noisy coherent states, generated from pure coherent states through the action of Gaussian additive noise.   Pure coherent states  have been successfully used in modeling a large number of physical systems \cite{coherentrmp}, including  the electromagnetic field, vibrational modes of solids, atomic ensembles, nuclear spins in a quantum dot, and Bose-Einstein condensates.  They  are important in continuous variable quantum protocols \cite{rmp1,rmp2}, such as quantum key distribution  \cite{qkd1,qkd2,qkd3,qkd4,qkd5}, cloning \cite{fiurasek,braunstein2001clonecoh,nonGau-opt,andersen2005clonecoh,namiki2006optimal}, and  quantum teleportation \cite{CV-tele}. In all these applications, the coherent state amplitude is Gaussian-modulated, meaning that the coherent state  $|\alpha\>    =  \sum_n  \,  e^{-|\alpha|^2/2}  \,  \alpha^n  |n\>/\sqrt{n!}$ is generated with probability  
\begin{align}\label{prob} p_\lambda (d^2 \alpha) =  \lambda  e^{-\lambda |\alpha|^2} ~  \frac{ d^2 \alpha}\pi \, .
\end{align}  
  Upon the action of the Gaussian additive noise, pure  coherent states are transformed into  displaced thermal states.     Specifically, the input coherent state $|\alpha\>$   is transformed into the displaced thermal state   
  \begin{align}
  \rho_{\mu,\alpha}   =   \int \frac{d^2\beta}\pi \,  \mu \,  e^{-\mu |\beta|^2}  \, |\alpha+ \beta\>\<\alpha+  \beta| \, .
  \end{align} 
   In this paper we consider the scenario where $N$ input  modes are independently prepared  in the state $\rho_{\mu,\alpha}$,  a common situation in several experiments, as copies of the same input  coherent state can be generated with standard techniques of phase locking \cite{drever1983laser,wiseman1993feedback,wiseman-feedback-book}. Given the $N$ input modes, we will search for the physical process that produces  $M$ output modes in the best  possible approximation of the pure amplified state $|g \alpha\>^{\otimes M}$, where $g \ge 1$ is the amplifier's gain.   
 The general problem considered in this paper includes as special cases the problems of coherent state amplification ($M=N=1$, $g>1$, $\mu \to \infty$)  \cite{caves1982amplification,ralph-lund-nondeterministic,namiki2011fundamental,caves2012quantumlimits,chiribellaxie,caves2013quantumlimits,caves2014noise,namiki2015amplification,caves2016models,ferreyrol2010implementation,heralded-amp,usuga2010noise,noiseless-amplifier,kocsis2013heralded,bruno2013complete}, cloning ($M\ge N$, $g=1$, $\mu \to \infty$) \cite{cerf2000cloning,braunstein2001clonecoh,nonGau-opt,andersen2005clonecoh,namiki2006optimal}, purification of noisy coherent states ($M\le N$, $g=1$,  and $\mu<\infty$) \cite{puri-exp-coherent-state,puri-prob-coherent-state}.    The problem of joint amplification and purification is similar in spirit to the task of  superbroadcasting (corresponding to $M \ge N$, $g=1$,  $\mu< \infty$) \cite{superbroadcasting-2,superbroadcasting-3}, with the difference that superbroadcasting optimizes  the fidelity of the individual output modes, whereas in our problem we  will focus on the global  fidelity,  quantifying how much the output modes globally resemble $M$ copies of the target state $|g\alpha\>$.

We will consider both deterministic and probabilistic processes.  In the deterministic processes the output is generated with  unit probability,  while in the probabilistic processes there is a non-zero chance of discarding the output.   
Probabilistic processes are interesting in that they can achieve enhanced performances in a variety of quantum tasks, including amplification \cite{ralph-lund-nondeterministic,chiribellaxie,caves2013quantumlimits,namiki2015amplification,caves2016models,ferreyrol2010implementation,heralded-amp,usuga2010noise,noiseless-amplifier,kocsis2013heralded,bruno2013complete}, cloning \cite{fiurasek}, and estimation \cite{fiurasek-probabilistic-estimation}.  While the probability of success can sometimes be unrealistically small, probabilistic processes are conceptually important because they provide the ultimate limits of what is possible in quantum mechanics. 

The study of probabilistic protocols is even more relevant when it comes to evaluating realistic experiments of quantum amplification. In this scenario, the key question is whether the experiment conclusively demonstrates the use of quantum resources, such as entanglement or coherence. More specifically, one wants to know whether the results of the experiment could be simulated by measuring the input systems, performing a classical computation, and generating the output systems in a quantum state that depends on the outcome of the computation. Startegies of this kind are called {\em measure-and-prepare protocols (M$ \&$P)}, {\em entanglement-breaking channels}, or also {\em classical strategies}. The performance of the best M\&P channel is the benchmark that needs to be surpassed in order to claim a genuine quantum processing. In this scenario, probabilistic M\&P protocols provide the strictest criterion of quantumness, because they certify that the results of the experiment could not be simulated without quantum resources, even with arbitrarily small probability.        
  
In this paper, we  provide the complete solution to the problem of optimal amplification of noisy coherent states, identifying both the optimal quantum strategies and the corresponding quantum benchmarks.  As an optimality criterion, we adopt the fidelity between the output and the ideal target of the amplification process.  It is worth stressing that the target output is a pure state, and therefore the fidelity has a direct operational interpretation as the probability of passing a test set up by a verifier \cite{braunstein2000criteria,yang2014certifying}.  The highest fidelity achievable by any M\&P protocol is called the {\em classical fidelity threshold (CFT)}.  Classical fidelity thresholds have been extensively studied for processes involving pure states  \cite{hammerer-prl-2005,adesso-chiribella2008,polzik-squeezingthelimit,namiki2008fidelity,bagan2009benchmark,namiki2011simple,chiribellaxie,chiribellaadesso2014,namiki2015quantum}. However, very little is known about the case where the states are mixed and the existing results are limited to the teleportation and storage of single-mode quantum states  \cite{polzik-squeezingthelimit,adesso-chiribella2008}. In this paper, we derive the quantum benchmarks for amplification-purification of displaced thermal states, along with the complete characterization of the optimal quantum strategies, both in the deterministic and in the probabilistic setting. 

The paper is structured as follows:  in Section \ref{sec:amplipuri}   we   formulate the problem of joint amplification and purification of noisy coherent states, giving a reduction to a single-mode problem.        When the single-mode problem involves amplification,   the optimal deterministic strategy is presented in  Section \ref{sec:optdet}, while the optimal probabilistic strategy is presented in Section \ref{sec:probopt}.    When the single-mode problem does not involve amplification,  we show that deterministic and probabilistic strategies perform equally well (Section  \ref{sec:puri}).    The quantum benchmarks for amplification and purification of noisy coherent states  are provided in  Section \ref{sec:benchmark}. Finally, the conclusions are drawn in Section \ref{sec:conclusion}.

\section{Formulation of the problem}\label{sec:amplipuri}

Here we consider the joint  amplification and purification of noisy coherent states by means of deterministic process.   Our goal is  to identify  the best quantum channel  that maps the input state $\rho_{\mu,\alpha}^{\otimes N}$ into the target state $|g\alpha\>^{\otimes M}$,  where the coherent amplitude $\alpha$ is Gaussian-distributed.   

\subsection{Figure of merit} 
As a figure of merit, we adopt the {\em global fidelity} between the output  of the channel and the target state. In formula, 
\begin{align}\label{globalfid}
F^{\rm det}_{N\to M}  (\alpha)   =  \<g\alpha |^{\otimes M}\map C(\rho_{\mu,\alpha}) |g\alpha\>^{\otimes M} \, ,
\end{align}  
where $\map C$ is a quantum channel (completely positive trace preserving map), sending states on the input Hilbert space $\map H _{in}  =  \spc H^{\otimes N}$ to the output Hilbert space $\map H _{out }  =  \spc H^{\otimes M}$, $\spc H$ being the Hilbert space associated to each mode. The map $\map C$ describes the input-output transformation occurring in the amplification process.     

Operationally, the global fidelity is  the probability of passing a test set up by a verifier who {\em (i)} knows the value of $\alpha$, and {\em (ii)} has access to the $M$ output modes.      
 Alternative choices of figure of merit are the \emph{single-copy fidelity} \cite{keyl1999optimal}---corresponding to the probability of passing a test where the verifier knows $\alpha$ but has  access a single output mode---or the trace distance \cite{nielsen2002quantum}---corresponding to the probability of passing a test where the verifier tries to distinguish between the channel output $\map C  \left(  \rho_{\mu,\alpha}^{\otimes N}  \right)$ and the target state  $|g\alpha\>^{\otimes M}$.   Note that operational interpretation of  the trace distance  presupposes that the verifier knows the channel $\map C$, in addition to the value of $\alpha$.   In the context of this paper, the fidelity is a more convenient choice  because it can be used to define benchmarks in the scenario where the channel $\map C$ is unknown.     

To identify the optimal quantum channel, we will be to find the channel $\map C$ that maximizes the average fidelity
\begin{align}\label{Fmultimode}
F_{N\to M,g}^{\rm det}=\int \frac{d^2 \alpha} \pi \,\, p_{\lambda}(\alpha) ~\<g\alpha |^{\otimes M}  \,  \map C\left(\rho_{\mu,\alpha}^{\otimes N}\right)  \, |g\alpha\>^{\otimes M} \, .
\end{align}
Due to the symmetry of the problem, the average fidelity includes as a special case the \emph{worst-case fidelity}, which can be obtained in the limit $\lambda \to 0$.   

When carrying out the optimization, we will make no assumptions on the channel  $\map C$. In particular, we will not assume that $\map C$  is Gaussian or covariant.   The only requirement---implicit in the fact that $\map C$ is trace-preserving---is that the amplification process happens \emph{\rm deterministically}, meaning that an output is produced with unit probability.  

\subsection{Reduction to  single-mode}
The first step towards the solution of the problem is the reduction to the single-mode scenario $M=N=1$.
The reduction is implemented  by   invertible transformations on the input and the output: specifically, one has   
\begin{align}
|g \alpha  \>^{\otimes M}
 \quad   \overset{U}{\longrightarrow}   \quad   &  |g \sqrt M \alpha  \>\otimes |0\>^{\otimes (M-1)} \\
\label{multimodemixed} \rho_{\mu, \alpha}^{\otimes N} \quad   \overset{U  \cdot U^\dag}{\longrightarrow}   \quad   &   \rho_{\mu,\sqrt N \, \alpha}  \otimes \rho_{\mu}^{\otimes {N-1}} 
 \, ,
\end{align} 
where $U$ is the Fourier  transform of  the modes, implementable through a network of beamsplitters \cite{reck1994experimental}.  Through this  mapping, the multimode is converted into a single-mode problem,  with the substitutions 
\begin{align}\label{substitutions}
g \to     g'    =   g  \sqrt{M/N}    \qquad {\rm and}  \qquad \lambda \to  \lambda'  =  \lambda/N        \, ,
\end{align} 
to be made in the target state for the process and in the probability distribution (\ref{prob}), respectively.   
The multimode fidelity (\ref{Fmultimode}) is then reduced to the single-mode fidelity 
\begin{align}\label{Fsinglemode}
F_{g'}^{\rm det}=\int \frac{d^2 \alpha} \pi \,\, p_{\lambda'}(\alpha) ~\<g'\alpha | \, \map C'\left(\rho_{\mu,\alpha}\right) \, |g'\alpha\> \, ,
\end{align}  
where $\map C'$ is the quantum channel implementing the reduced, single-mode transformation.  

At this point, the problem is to find the process that maximizes the single-mode fidelity (\ref{Fsinglemode}).  A convenient approach is the semidefinite programming method developed in  Ref.  \cite{chiribellaxie}, which gives an explicit expression for the optimal fidelity, denoted by $F^{\rm det*}_{g'}$. Precisely, we have 
\begin{align}\label{1}
F^{\rm det*}_{g'} = \inf_{\sigma}  \left\| \int  \frac{\d^2 \alpha}{\pi} p_{\lambda'} (\alpha)    |g' \alpha\>\<  g'  \alpha|  \otimes    \sigma^{- \frac12}  \rho_{\mu,\overline \alpha}    \sigma^{-\frac 12}\right\|_{\infty}  \, ,
\end{align}
where $\sigma$ is arbitrary quantum state of  the input mode space and  $\|  A     \|_{\infty}  =  \sup_{\|  |\psi\> \|=1}  \,  \|  A \,  |\psi\>\|$  is the operator norm of the operator $A$.

\section{Optimal deterministic processes: the $g'\ge 1$ case}\label{sec:optdet}

For $g'\ge 1$,  explicit calculation of the optimal deterministic fidelity (\ref{1}) gives the value
\begin{align}\label{det-opt}
F^{\rm det*}_{g'\ge 1 }&=\begin{cases}
\dfrac{N_{\rm C}+N_{\rm T}+1}{g'^2N_{\rm C}(N_{\rm T}+1)},& g'\ge \frac{N_{\rm C}+N_{\rm T}+1}{N_{\rm C}}  \\ 
&\\
\frac{1}{(g'-1)^2N_{\rm C}+N_{\rm T}+1},& g'< \frac{N_{\rm C}+N_{\rm T}+1}{N_{\rm C}}  .
\end{cases}
\end{align}
where  $N_{\rm C}=  1/{\lambda'}  =  N/\lambda$ is the expected  number of  photons in the signal and $N_{\rm T}=1/\mu$ is  the expected number of photons added by thermal noise.   The details of the derivation are presented  in Appendices  \ref{app:optimalfidelity} and  \ref{app:twomodesqueeze}. 

The optimal fidelity can be attained with standard quantum optics techniques. For the single-mode problem, the optimal deterministic strategy  is to couple the input mode with an ancillary mode, initially in the vacuum, so that the two modes undergo a two-mode squeezing transformation.    Eventually, the ancillary mode is discarded.  Mathematically, this sequence of operations    is described by the quantum channel   
\begin{align} \mathcal{C}_r(\rho)=\mathrm{Tr}_{B} \left[e^{r(a^{\dag}b^{\dag}-ab)}(\rho\otimes|0\rangle\langle0|)e^{-r(a^{\dag}b^{\dag}-ab)} \right] \, ,\end{align} 
where $a$ and $b$ are the annihilation operators of the input mode and the ancillary mode, respectively, and $\Tr_B$ is the partial trace over the ancillary mode. 
In order to achieve the maximum fidelity the  squeezing parameter $r$ must be tuned to satisfy the condition 
\begin{align}\label{cosh}\cosh r =\frac{g' N_{\rm C}}{1+N_{\rm C}+N_{\rm T}}
\end{align} when $g'\geq (N_{\rm C}+N_{\rm T}+1)/N_{\rm C}$ and
 $r=0$ otherwise (see Appendix \ref{app:twomodesqueeze}). 
    Note that the choice  $r=0$ corresponds to the identity channel: when the expected number of thermal photons is increased,  the optimal strategy is just  to leave the input state untouched.   In the multimode scenario, this means that  the best amplification setup  for high-temperature states consists of  a network of   beamsplitters. 
 
 \medskip 
 
{\bf Remark 1 (Gaussianity vs non-Gaussianity).}  We have seen that the optimal deterministic amplification is achieved by Gaussian operations. It is worth stressing that Gaussianity was not assumed from the start, but came as a result of  the optimization.  This result  may seem to be in  contrast with the earlier work by Cerf \emph{et al} \cite{nonGau-opt}, who showed that Gaussian operations are suboptimal for the problem of cloning coherent states, corresponding to $N=1$, $M=2$, $g=1$, and $\mu  \to  \infty$.    The reason for the discrepancy is that  Wolf \emph{et al} focussed on the \emph{local} fidelity---that is, the fidelity of each individual clone with the target state. By employing non-Gaussian operations,  their protocol manages to produces clones that better resemble the target state, when examined individually. Still, when the two clones are examined jointly, the optimal cloning operation is Gaussian and coincides with the cloner proposed by Cerf, Ipe, and Rottenberg  \cite{cerf2000cloning}  (see also \cite{braunstein2001clonecoh}). 

\medskip 
 
 {\bf Remark 2  (amplification vs purification).}   For   $g'\ge 1$,         
 the output states of the optimal single-mode process are more mixed than the input states.  Indeed,  the number of thermal photons goes from $N_{\rm T}= 1/\mu$ in the input state to  $N_{\rm T}'   =\cosh^2 r \, N_{\rm T}  + (\cosh^2 r -1)$ in the output state.    Since $N_{\rm T}'$ is larger than $N_{\rm T}$,  no  purification takes place.   Summarizing:    for $g'\ge 1$,   amplification dominates over purification in the single-mode setting.  Let us look at the multimode scenario.        There, the total  number of thermal photons in the input (summing the contributions from all modes) is $  N_{\rm total}=     N/\mu $. The  total number of thermal photons in the output  is 
 \begin{align}
 \nonumber N_{\rm total}'     &=\cosh^2 r \, \frac {N_{\rm total}}{N}  + (\cosh^2 r -1) \\
    &=     \left(  \frac{g N_{\rm C}}{1+N_{\rm C}+N_{\rm total}/N}\right)^2     \frac {  M  \,  }{N}    \left(  \frac{N_{\rm total}}N     +1\right)   - 1   \, ,
 \end{align} 
 having used Eq. (\ref{cosh}).   The relation can also be expressed in terms of individual input and output modes as  
 \begin{align}
 \nonumber N_{\rm single}'     
    &=     \left(  \frac{g N_{\rm C}}{1+N_{\rm C}+N_{\rm single}}\right)^2     \frac {  1 \,  }{N}    \left(  {N_{\rm single}}     +1\right)   - \frac 1M   \, ,
 \end{align} 
 with $N_{\rm single} =  N_{\rm total}/N$ and $N_{\rm single}' =  N_{\rm total}'/M$. 
 The above equation quantifies the competition between amplification and purification for deterministic processes and for $   g  \ge    \sqrt {N/M}$.  

\section{Optimal  probabilistic processes: the $g'\ge 1$ case}  \label{sec:probopt}

Non-deterministic processes are known to boost the performances of amplification   \cite{ralph-lund-nondeterministic,chiribellaxie}, a mechanism that has been observed experimentally in the case of pure states  \cite{heralded-amp}. Still, the case of mixed states has remained unexplored so far. To tackle the problem, we model a generic non-deterministic amplification process by a trace non-increasing completely positive map  $\map Q$, which describes the occurrence of a desired transformation heralded by a suitable measurement outcome. 
    In this setting, the fidelity is given  by
\[F_{N\to M,g}^{\rm prob}=\frac{\int \frac {d^2 \alpha }\pi\,\, p_{\lambda}  (\alpha)~ \<g\alpha |^{\otimes M}\map Q \left(\rho^{\otimes N}_{\mu,\alpha} \right) |g\alpha\>^{\otimes M}}{\int \frac{d^2 \alpha } \pi\,\, p_{\lambda}(\alpha)~ \Tr\left[ \map Q\left(\rho^{\otimes N}_{\mu,\alpha} \right) \right] } \, .\] 
In general, the  fidelity depends on the probability of the heralded outcome \cite{caves2013quantumlimits}. Here we will allow the probability to be arbitrarily small, thus giving the ultimate quantum fidelity achievable by arbitrary probabilistic processes.    

The problem can be reduced to a single-mode problem, as illustrated in the deterministic case.   Using the technique of   \cite{fiurasek,chiribellaxie}, the ultimate fidelity for probabilistic amplification can be expressed as  
\begin{align}\label{prob_opt_gen}
F^{{\rm prob}*}_{g' } =   \left\| \int  \frac{\d^2 \alpha}{\pi} p_{{\lambda'}} (\alpha)    |g' \alpha\>\<  g'  \alpha|  \otimes    \<  \tilde \rho\>^{- \frac12}  \rho_{\mu,\overline \alpha}    \<\tilde \rho\>^{-\frac 12}\right\|_{\infty}
\end{align}
  where   $\<\tilde \rho\>$ is the average state of the  ensemble  $\{\rho_{\mu,\overline \alpha} \, ,  p_{\lambda'} (\alpha)\}$.    
  
By explicit calculation (Appendix \ref{app:optprob}), we obtain  the ultimate probabilistic fidelity   
\begin{equation}\label{prob-opt}
F^{{\rm prob}*}_{g'\ge 1}=
\begin{cases}
\dfrac{N_{\rm C}+N_{\rm T}+1}{g'^2N_{\rm C}(N_{\rm T}+1)},& g'\ge\frac{\sqrt{(N_{\rm C}+N_{\rm T}+1)(N_{\rm C}+N_{\rm T})}}{N_{\rm C}}\\
&  \\ 
\frac{N_{\rm C}+N_{\rm T}}{N_{\rm C}+N_{\rm T}+g'^2N_{\rm C}N_{\rm T}},& g' <\frac{\sqrt{(N_{\rm C}+N_{\rm T}+1)(N_{\rm C}+N_{\rm T})}}{N_{\rm C}}   \, .
\end{cases}
\end{equation}
Note that in  the region $g' \ge (N_{\rm C}+N_{\rm T}+1)/N_{\rm C}$, there is no difference between the maximum probabilistic fidelity and the maximum deterministic fidelity in  Eqs.(\ref{det-opt}) and (\ref{prob-opt}), respectively. This means that arbitrary probabilistic setups with arbitrary success probability cannot do better than the best deterministic setup.  A similar phenomenon occurs in the amplification of Gaussian-distributed coherent states  \cite{chiribellaxie}, where the advantage of probabilistic setup disappears after the amplification parameter exceeds a critical threshold.  

For  values of $g' $ between $1$ and   $ (N_{\rm C}+N_{\rm T}+1)/N_{\rm C}$  there is a gap between the performance of probabilistic and deterministic processes.  For example, when the input states are pure $(N_{\rm T}   =  0)$, we obtain  $F^{{\rm prob}*}_{g'\ge 1}  =  1$, in agreement with the existence of noiseless probabilistic amplifiers   \cite{ralph-lund-nondeterministic,chiribellaxie}.  
 An interesting question is whether  noiseless amplification is possible for mixed states $(N_{\rm T}    >  0)$.   Our result answers the question in the negative, showing that the probabilistic fidelity is never equal to 1, except in the trivial case where the input state is perfectly known ($N_{\rm C}  =  0$).

The ultimate probabilistic fidelity can be achieved   by a non-deterministic noiseless amplifier of the kind proposed by Ralph and Lund  \cite{ralph-lund-nondeterministic,heralded-amp}. Mathematically, the non-deterministic amplifier is described by the map 
\begin{align}\label{Q}
Q_{K}(\rho)=Q_{K}\rho Q_{K}^{\dagger}\, , \qquad  Q_{K}  =   y^{-K}  \, \sum^{K}_{n=0}y^{n}|n\rangle\langle n| \, ,
\end{align}  
where  $K$ is a large integer (ideally approaching  infinity) and  $y$ is a suitable parameter depending on the desired degree of amplification. 

In order to approach  the ultimate probabilistic fidelity  (\ref{prob-opt}), the amplification parameter $y$ has to be  tuned as 
\begin{align}
y=\frac{g'N_{\rm C}}{N_{\rm C}+N_{\rm T}}  \, ,
\end{align}
for 
values of $ g' $ between  ${\sqrt{(N_{\rm C}+N_{\rm T}+1)(N_{\rm C}+N_{\rm T})}}/{N_{\rm C}} $ and $1$,   
and as
\begin{align}  
 y=\frac{N_{\rm C}+N_{\rm T}+1}{g'N_{\rm C}}
 \end{align}
 for values of   $g'$ between  ${\sqrt{(N_{\rm C}+N_{\rm T}+1)(N_{\rm C}+N_{\rm T})}}/{N_{\rm C}} $ and $ {(N_{\rm C}+N_{\rm T}+1)}/{N_{\rm C}}$. 
 Choosing the above values, the  fidelity of the non-deterministic amplifier (\ref{Q}) becomes exponentially close to the optimal probabilistic fidelity in the large $K$ limit  (cf. Appendix \ref{app:ralphflund}).

\medskip

 {\bf Remark (amplification vs purification)}.  For $g'\ge 1$, the output states of the optimal process are more mixed then the input states, also in the probabilistic setting.   Let us look at  the region $1\le g'\le(N_{\rm C}+N_{\rm T}+1)/N_{\rm C}$,  where the probabilistic advantages are more prominent.   Here the expected number  of thermal photons is 
 \begin{align}   N'_{\rm T }=  N_{\rm T}  \,      \, \frac{  y^2   \, \mu}{1+\mu-y^2} \, ,
 \end{align}  
 which cannot be smaller than  $N_{\rm T}$, since the amplification parameter $y$ is  larger than or equal to $1$.      In summary, no purification takes place at the single-mode level.   Again, the situation is more nuanced in the multimode case,  
  where the number of thermal photons is initially larger by a factor $N$.   
  Explicitly, the total number of thermal photons, initially equal to $N_{\rm total}=  N/\mu$, is finally equal to 
 \begin{align}   N'_{\rm total }=  \frac{ N_{\rm total} ~   M\,    \mu   g^2   N^2_{\rm C}    }{ N^2  \left[ (1+\mu)  \left(N_{\rm C}+\frac{N_{\rm total}}{N}\right)^2-  (g   N_{\rm C})^2  \frac M N  \right]} \, . 
 \end{align}   
Equivalently, the number of thermal photons per mode
goes from $N_{\rm single}=  1/\mu$ to 
\begin{align}   N'_{\rm single }=       \,     \, \frac{   N_{\rm single}~   \mu \,  g^2 \,   N^2_{\rm C}  }{ N  \left[  (1+\mu)  (N_{\rm C}+N_{\rm single})^2-  (g   N_{\rm C})^2  \frac MN \right]} \, . 
 \end{align}   
According the the above equation, the values of the parameter determine whether purification can  take place in conjunction with amplification.  

\section{The $g' \le 1$ case: no advantage from probabilistic operations}\label{sec:puri}
 Let us consider now the case where $g' \le  1$.     In the single-mode picture, the task is to transform a mixed input state into a purer, albeit attenuated, output state.  
  Quite surprisingly,  we find that in this case there is no difference between the optimal performance of deterministic and probabilistic processes. Specifically, the exact calculation of the optimal fidelities yields the value     
  \begin{align}\label{puri}
F^{\rm det*}_{g'\le 1}=F^{{\rm prob}*}_{g'\le 1}=
\frac{N_{\rm C}+N_{\rm T}}{N_{\rm C}+N_{\rm T}+g'^2N_{\rm C}N_{\rm T}} \, .
\end{align}
(see Appendix \ref{app:optimalfidelity} for the details).   Eq. (\ref{puri}) tells us that   there is no fidelity-probability tradeoff in the purification regime: the fidelity has the same value for every value of the success probability.  Even if we postselect on extremely rare events, these events cannot increase the performance of our purification setup.   This situation has to be contrasted with the case of amplification, where the reduction of the success probability is accompanied by an increase in fidelity.   

Note that, the fidelity formula (\ref{puri})  can be applied to the special case of  purification of $N=2$ noisy coherent states, corresponding to $g=1$ and $g'=1/\sqrt 2$.   In the limit of infinite modulation $N_{\rm C}  \to \infty$, we retrieve the  fidelity from the earlier work of Andersen {\em et al}  \cite{puri-exp-coherent-state}.

The ultimate quantum fidelity of Eq.(\ref{puri}) can be attained via  the attenuation channel 
\begin{align}
\map C_\theta (\rho) = \Tr_B \left[e^{i\theta(a^\dag b-b^\dag a)}(\rho\otimes |0\>\<0|) e^{-i\theta(a^\dag b-b^\dag a)}\right]  \, ,
\end{align} 
where the angle $\theta$ has to be adjusted to satisfy the condition 
\begin{align}
\cos  \theta=\frac{g'}{1+{N_{\rm T}}/{N_{\rm C}}}  \,  .
\end{align}

By definition,  no amplification takes place at the single-mode level for $g'< 1$.  
    The situation is different at the multimode level: for $g <  N/M$ it is still possible to have amplification ($ g>1$), provided that $N$ is larger than $M$.   In this setting, some of the input modes are sacrificed, in order to allow for the joint amplification and purification of the output modes.     Note that  postselection and other probabilistic operations do not contribute to the tradeoff:  the best way to jointly amplify and purify  noisy coherent states is deterministic. 


\section{Quantum benchmark}\label{sec:benchmark}

We identified the optimal setups for the joint amplification and purification of noisy coherent states, both in the deterministic and in the probabilistic scenario.   The results obtained so far  are appealing because the optimal quantum processes  can be achieved using present-day technology.  Still, real experiments are typically subjects to imperfections and, as a result, the optimal performance may not be exactly attained.    The question is how to certify that the experiment could not be reproduced with  classical (a.k.a. M\&P) strategies, by just estimating the input state and, based on the outcome, preparing the output state.  In order to rule out this possibility, it is important to know the value of the classical fidelity threshold, which provides the benchmark that has to be passed in order to claim a successful implementation of quantum amplification. 

We will start from the CFT corresponding to probabilistic M\&P strategies, where one is allowed to discard unfavourable measurement outcomes.  By considering strategies with arbitrary probability of success, we will obtain the most stringent benchmark one can choose.  
In the single-mode scenario, the probabilistic CFT can be computed using the method of  \cite{chiribellaxie}, as   
 \begin{align}
F^{c}_{g'} =   \left\| \int  \frac{\d^2 \alpha}{\pi} p_{{\lambda'}} (\alpha)    |g \alpha\>\<  g  \alpha|  \otimes    \<  \tilde \rho\>^{- \frac12}  \rho_{\mu,\overline \alpha}    \<\tilde \rho\>^{-\frac 12}\right\|_{\times}
\end{align}
  where    where   $\<\tilde \rho\>$ is the average state of the  ensemble  $\{\rho_{\mu,\bar \alpha} $  and 
  \begin{align}
  \| A\|_\times  =  \sup_{ \| |\varphi \>  \|   = \|  |\psi \> \|=  \| | \varphi'\>\|   =  \| |\psi'\> \| = 1}  \,  |    \<\varphi| \<\psi|  A   |\varphi'\>  |\psi'\> |
  \end{align} is the injective cross norm.   

By evaluating the norm  (Appendix \ref{app:CFT}), we find  the probabilistic CFT 
\begin{align}\label{cft explic n}
&F_{g'}^{c}=\dfrac{N_{\rm C}+N_{\rm T}'}{N_{\rm C}+\widetilde N_{\rm T}+N_{\rm C} \widetilde N_{\rm T}g'^{2}} \, ,
\end{align}
with  $\widetilde N_{\rm T}=N_{\rm T}+1$.  Note  that the CFT is always lower than the quantum limits established for both quantum deterministic and probabilistic process. This means that, in principle, there is always a way to certify the quantumness of a realistic setup for amplification.

Quite remarkably, we find that  the probabilistic CFT   can be achieved deterministically by performing the heterodyne measurement with POVM elements  $P(\beta)  =  |\beta\>\<\beta|$ and conditionally re-preparing the pure coherent state $|z\beta\>$, with  $z=g'N_{\rm C}/(N_{\rm C}+N_{\rm T}+1)$ (see Appendix \ref{app:CFT}).  In other words, postselection is completely useless when searching for a classical strategy for amplification and purification of noisy coherent states: no matter how small the probability of success is, probabilistic M\&P channels cannot  do better than the optimal deterministic channel.

\medskip

\section{conclusions}\label{sec:conclusion}

In this paper we investigated a general scenario that interpolates between the tasks of amplification and purification. We identified  the optimal physical process that produces the best approximation of a pure and amplified coherent state  from multiple copies of a Gaussian-distributed noisy coherent state.   We carried out an \emph{ab initio} optimization both for deterministic processes and for probabilistic processes, showing how to implement the optimal processes using existing techniques in quantum optics. Specifically, the optimal deterministic process can be implemented using a network of beamsplitters and a two-mode squeezing operation, while the optimal probabilistic process uses the nondeterministic amplifier by Ralph and Lund  \cite{ralph-lund-nondeterministic}, again, combined with a network of beamsplitters. 

We proved that probabilistic operations outperform their deterministic counterpart in a certain region of the parameter space.  However, there is also a parameter region where using probabilistic operations  offers no advantage, irrespectively of the probability.  In fact, there exists even a region where the best amplification scheme consists in a passive optical network, consisting only of beamsplitters. 

Since all the optimal protocols identified in our work are experimentally feasible, it is important to have criteria to witness quantum advantages over classical amplification techniques. In the paper we provided rigorous benchmarks that can be used to establish such advantages.   Remarkably,  the value of the benchmark is independent of the required probability of success: classical deterministic strategies and classical strategies using postselection perform equally well.   It is also worth noting that the value of the benchmark is strictly smaller than the optimal quantum fidelity for every value of the parameters.  This result establishes that the joint amplification and purification of noisy coherent states is a genuinely quantum task.  It is our hope that this work will stimulate the realization of new experiments and the progress in the implementation of optimized optical setups  that approach the ultimate quantum limit.

\medskip

{\bf Acknowledgements.} This work is supported by the Hong Kong Research Grant Council
through Grant No. 17326616, by National Science
Foundation of China through Grant No. 11675136, 
by the HKU Seed Funding for Basic Research, and by  the Canadian Institute for Advanced Research. 
\bibliographystyle{apsrev4-1}
\bibliography{GaussianAmplification}

\appendix

\section{Proof that the r.h.s. of Eq. (\ref{det-opt}) is an upper bound to the  deterministic fidelity} \label{app:optimalfidelity}

Our goal is to evaluate the operator norm  of the operator  
\begin{align}
\Gamma_\sigma  = \int  \frac{\d^2 \alpha}{\pi} p_{\lambda'} (\alpha)    |g' \alpha\>\<  g'  \alpha|  \otimes    \sigma^{- \frac12}  \rho_{\mu,\overline \alpha}    \sigma^{-\frac 12}
\end{align}
 and  to minimize the norm over the state $\sigma$.       As an ansatz, we choose  $\sigma$ to be a thermal state, of the form 
\begin{align*}
& \sigma_\kappa=\int \frac{d^2 \alpha}{\pi} \kappa e^{-\kappa |\alpha|^2}|\alpha\>\<\alpha|\, ,   \qquad \kappa\ge 0 \, .
\end{align*}

This choice gives us an upper bound of the optimal fidelity, as 
\begin{align}\label{upperbound}
F^{\rm det*}_{g'}&=\|\Gamma_{\sigma}\|_\infty\le\| \Gamma_{ \sigma_\kappa}\|_\infty \, .
\end{align}
Now, the operator  norm is given by $\| \Gamma_{ \sigma_\kappa}\|_\infty=\lim_{p\to \infty}(\Tr | \Gamma_{\sigma_\kappa}|^p)^{1/p}$. Calculating  the trace  we obtain 
\begin{align}\label{b1}
  & \Tr \left[  \Gamma_{\sigma_\kappa}^p\right]\\
&\nonumber =   \left[ \frac{\mu {\lambda'}(\kappa+1)}{\kappa}\right]^p  \,  \int  \frac{\d^{2p}  \bs \alpha}{\pi^p}  \int   \frac{\d^{2p}  \bs \beta}{\pi^p}  \,   e^{  -      (\bs  \alpha  \oplus  \bs \beta)^\dag   M_p   (\bs \alpha  \oplus \bs \beta)}  \, ,
\end{align}
where   $\bs \alpha \in  \C^p$  and $  \bs \beta  \in  \C^p$ are the column vectors defined by $\bs \alpha  =   (\alpha_1,\dots,  \alpha_p)^T$ and $\bs \beta  =   (\beta_1,\dots,  \beta_p)^T$, respectively,   $\bs \alpha  \oplus  \bs \beta  \in\C^{2p}$ is the column vector    $\bs \alpha  \oplus  \bs \beta   = (\alpha_1,\dots,  \alpha_p  ,   \beta_1,\dots,  \beta_p)^T$, and $M_p$ is the  $2p\times 2p$ matrix  defined by   $M_p   =   \begin{pmatrix}   A    &    B  \\
  B   &  C \end{pmatrix}$   with
  \begin{align*}
  A_{ij}    =&   ({\lambda'}  +1+g'^2)  \,  \delta_{ij}       -   g'^2  \delta_{  i, (i+1)  \mod  p} \\& -  (\kappa+1)    \,  \delta_{  i, (i-1)\mod p}  \\
  B_{ij}    =&      \delta_{ij}       -      (\kappa+1)   \,    \delta_{  i, (i-1)  \mod p}      \\
  C_{ij}    =&    (\mu+ 1)   \delta_{ij}       -      (\kappa+1)   \,       \delta_{  i, (i-1)  \mod p}  \, .
  \end{align*}
 Note that  $A, B$, and $C$ are circulant matrices and, therefore, they are diagonal in the Fourier basis.   Hence, the matrix $M_p$ can be rewritten as    $M_p    =    U    M_p' U^\dag$, with $M_p'  =  \begin{pmatrix}    A'  &      B'  \\  B' &  C' \end{pmatrix} $,   where $A',  B',$ and $C'$ are diagonal matrices and $U$ is the block diagonal matrix  $U  =   \begin{pmatrix}    F &      \bs 0    \\  \bs   0 &  F \end{pmatrix}$,   $F$  being the Fourier transform.     Finally,  the matrix $M_p'$ can be expressed as  $M_p'  =    U_\pi    \widetilde M_p  U_{\pi}^\dag $, where  $U_\pi$ is a permutation matrix and $\widetilde M_p$ is a direct sum of  two-by-two matrices, which in turn can be diagonalized.     As a result, we obtain  the relation
 \begin{align}
  \nonumber \Tr \left[  \Gamma_{\sigma_\kappa}^p\right] & =   \left[ \frac{\mu {\lambda'}(\kappa+1)}{\kappa}\right]^p  \,     \frac  1 {  \det  \widetilde M_p}    \\
  \label{aaa} &  =  \left[ \frac{\mu {\lambda'}(\kappa+1)}{\kappa}\right]^p  \,     \frac  1 {  \det  M_p}   \, ,
   \end{align}
 where we used the fact that $\widetilde M_p$ and $M_p$ are unitarily  equivalent and therefore have the same determinant.
    Using Eq. (\ref{aaa}), we can compute the norm of $\widetilde \Gamma_{\sigma_y}$ as
\begin{align}\label{b3}
\|     \Gamma_{\sigma_\kappa}  \|_{\infty}     & =       \frac{\mu {\lambda'}(\kappa+1)}{\kappa}     \,      \frac 1 {  \lim_{p\to  \infty}  \,\left(\det  M_p  \right)^{1/p}}
\end{align}
  The determinant of $M_p$  can be computed with the relations $\det M_p =\det   (  AC-B^2)$  and
  \begin{align*}
  (AC-B^2)_{ij}    &=   a   \,  \delta_{ij}       - b \,   \delta_{  i, (i+1)  \mod  p}    -  c  \,  \delta_{  i, (i-1)  \mod p} \, .
  \end{align*}
  with \begin{align*}
a  &=   \mu+{\lambda'} +\mu{\lambda'} +g'^2(\mu+\kappa+2) \\
b  &  =   g'^2(\mu+1)    \\
c  &  =   (g'^2+\mu+{\lambda'})(\kappa+1)   \,  .
  \end{align*}
Since $AC-B^2$ is a circulant matrix, its eigenvalues are given by ${\lambda'}_{p,n}  =  a  -  b\, \omega_p^{-n} - c\, \omega_p^{n} $, where $\omega_p   =   \exp[ 2\pi  i/p]$. Hence, we have
 \begin{align*}
\lim_{p\rightarrow\infty}     \,  \ln \left(  \det M_p \right )^{1/p}   & =\lim_{p\rightarrow\infty}  \, \frac{1}{p}  \, \sum^{p-1}_{n=0}  \,   \ln    {\lambda'}_{p,n}\\
 & =\lim_{p\rightarrow\infty}  \,  \frac{1}{p}\sum^{p-1}_{n=0}   \,  \ln(a -b \, \omega_p^{-n} - c \, \omega_p^{n})\\
&=\int^{2\pi}_0  \, \frac{d\theta}{2\pi}  \,   \ln[b(y_+-e^{-i\theta})(1-y_-e^{i\theta})] \, ,
\end{align*}
with  $y_\pm=\frac{a\pm \sqrt{a^2-4bc}}{2b}$. 

Now, choosing $\kappa \leq \frac{{\lambda'}\mu}{{\lambda'}+\mu}$ we can satisfy the conditions $y_+\geq1$ and $y_-\leq1$.   Under these conditions, the above  integral can be evaluated, giving the value  $\ln(b y_+)$. Hence, the upper bound  (\ref{upperbound})   becomes 
\begin{align}\label{eq:a5}
\mathrm{F}_{g'\ge 1}^{\rm det*}&\le\frac{\mu{\lambda'}(\kappa+1)}{\kappa}\frac{2}{a+\sqrt{a^2-4bc}} \, .
\end{align}

By minimizing over $\kappa$,  we  obtain the optimal upper bound.  The optimal value of $\kappa$  is 
\begin{align*}
\kappa^*=  \left\{ 
\begin{array}{ll}   \frac{{\lambda'}\mu}{{\lambda'}+\mu}  \qquad   & g'\ge ({\lambda'}+\mu+{\lambda'}\mu)/\mu  \\
  &  \\ 
  \frac{\mu(g'-1)^2+{\lambda'}(\mu+1)}{g'(g'+\mu)} \qquad & g'< ({\lambda'}+\mu+{\lambda'}\mu)/\mu \, .
  \end{array}  
  \right. 
  \end{align*}
Inserting the above values in  Eq.(\ref{eq:a5})  we obtain the r.h.s. of Eq. (\ref{det-opt}).   \qed

\section{Proof of  Eq. (\ref{prob-opt}) }\label{app:optprob}
 The evaluation of the probabilistic fidelity follows the same steps used in the previous section.     The only difference is that we have to fix the state $\sigma$ to $\sigma  =  \<\tilde \rho \>$,  the  (conjugate of) average state of the input ensemble.     

The state $\<\tilde \rho  \> $ is a thermal state, of the form  $\sigma_{\kappa'}$ with  $\kappa'=\frac{{\lambda'}\mu}{{\lambda'}+\mu}$. By substituting the the value of $\kappa'$ into Eq.(\ref{b3}) and using Eq.(\ref{prob_opt_gen}), we  get the ultimate fidelity achievable for arbitrary probabilistic processes, in the form of  in Eq.(\ref{prob-opt}).  \qed

\section{Optimality of  two-mode squeezing}\label{app:twomodesqueeze}

For the parametric  amplifier channel  
\begin{align} \map{C}_r(\rho)=\mathrm{Tr}_{B}\left[e^{r(a^{\dag}b^{\dag}-ab)} \, (\rho\otimes|0\rangle\langle0| ) \, e^{-r(a^{\dag}b^{\dag}-ab)}\right] \, ,
\end{align}
 the fidelity is
 \begin{align}
F^{r}_{g'\ge 1}=\dfrac{{\lambda'}\mu}{{\lambda'}(\mu+1)\cosh^{2}r+\mu(g'-\cosh r)^{2}} \, .
\end{align}

By optimizing over $r$,   we get the maximum value  advertised in  Eq.(\ref{det-opt}). The maximum is attained by choosing $r$ such that  $\cosh r =g' N_{\rm C}/(1+N_{\rm C}+N_{\rm T})$, in the case $g'\geq (N_{\rm C}+N_{\rm T}+1)/N_{\rm C}$, and by choosing $\cosh r=1$ otherwise.

\section{\label{sec:level1} Optimality of the noiseless nondeterministic amplifier}\label{app:ralphflund}

The noiseless amplifier is described by the  quantum operation $Q_{K}(\rho)=Q_{K}\rho Q_{K}^{\dagger}$ with $Q_{K}\propto\sum^{K}_{n=0}y^{n}|n\rangle\langle n|$. Its  fidelity is given by
\begin{widetext}
\begin{equation}\begin{split}
&F_{g'\ge 1,K}^{\rm prob}=\dfrac{\int\dfrac{d^{2}\alpha}{\pi}{\lambda'} e^{-{\lambda'}|\alpha|^{2}}\langle g \alpha |Q_{K}\rho_{\alpha,x}Q_{K}^{\dagger}|d\alpha\rangle}{\int\dfrac{d^{2}\beta}{\pi}{\lambda'} e^{-{\lambda'}|\beta|^{2}}\mathrm{Tr}[\rho_{\beta,x}Q_{K}^{\dagger}Q_{K}]}\\
&=\dfrac{\int\dfrac{d^{2}\alpha}{\pi}\int\dfrac{d^{2}\gamma}{\pi} e^{-\mu|\gamma|^{2}} e^{-{\lambda'}|\alpha|^{2}}e^{-(1-y^{2})|\gamma+\alpha|^{2}}|\langle g' \alpha|P_{K}|y(\gamma+\alpha)\rangle|^{2}}{\int\dfrac{d^{2}\beta}{\pi}\int\dfrac{d^{2}\gamma}{\pi} e^{-\mu|\gamma|^{2}} e^{-{\lambda'}|\beta|^{2}}e^{-(1-y^{2})|\gamma+\beta|^{2}}\langle y(\gamma+\beta)|P_{K}|y(\gamma+\beta)\rangle}\\
& \geq C(y)\int\dfrac{d^{2}\alpha}{\pi}\int\dfrac{d^{2}\gamma}{\pi} e^{-\mu|\gamma|^{2}} e^{-{\lambda'}|\alpha|^{2}}e^{-(1-y^{2})|\gamma+\alpha|^{2}}[|\langle g' \alpha|y(\gamma+\alpha)\rangle|^{2}-2|\langle g' \alpha |(I-P_{K})|y(\gamma+\alpha)|]\\
&\geq \dfrac{({\lambda'}+\mu)(1-y^{2})+{\lambda'}\mu }{{\lambda'}+\mu+{\lambda'}\mu+g'^2-y^2g'^2+\mu g'^2-2yg'\mu}-2\sqrt{\mathbb{E}[\langle g' \alpha |(I-P_{K})|g'\alpha\rangle]\mathbb{E}[\langle y(\gamma+ \alpha )|(I-P_{K})|y(\gamma+\alpha)\rangle]}
\end{split}\end{equation}\end{widetext}
where  $\mathbb{E}(f)$ denotes the expectation value of the function $f$ over the Gaussian distribution
$p(\alpha,\gamma)=C(y)e^{-\mu|\gamma|^{2}}e^{-{\lambda'}|\alpha|^{2}}e^{-(1-y^{2})|\gamma+\alpha|^{2}}$, with $C(y)=({\lambda'}+\mu)(1-y^{2})+{\lambda'}\mu$.   

The expectation values can be computed explicitly as 
\begin{align*}
&\mathbb{E}[\langle g' \alpha |(I-P_{K})|g'\alpha\rangle]
\\&=\left[\dfrac{g'^{2}}{g'^{2}+{\lambda'}+1-y^{2}-\dfrac{(1-y^{2})^{2}}{\mu+1-y^{2}}}\right]^{K+1}\\
\end{align*}
and
\begin{align*}
&\mathbb{E}[\langle y(\gamma+ \alpha) |(I-P_{K})|y(\gamma+\alpha)\rangle]
\\&=\left[\dfrac{y^{2}}{1+{\lambda'} -\dfrac{{\lambda'}^{2}}{{\lambda'}+\mu}}\right]^{K+1}  \, .
\end{align*}
Note that both expectation values vanish exponentially fast in the large $K$ limit. 

Now, we tune the amplification parameter $y$ in order to attain the maximum fidelity: 
\begin{enumerate}
\item for  $g'$ between $1$ and $\sqrt{\frac{{\lambda'}^2}{\mu}+{\lambda'}+(1+\frac{\lambda'}\mu)^2}$,  we choose set  $y=\frac{g'\mu}{{\lambda'}+\mu}$,  obtaining fidelity $F^{\rm prob}_{K}\rightarrow\dfrac{{\lambda'} +\mu}{{\lambda'} +\mu+ g'^{2}}$ in the limit $K\to\infty$. 
\item For $g'$ between $ \sqrt{\frac{{\lambda'}^2}{\mu}+{\lambda'}+(1+\frac{\lambda'}\mu)^2}$ and $\frac{{\lambda'}}{\mu}+{\lambda'}+1  $,   we set $y=\frac{{\lambda'}+\mu+{\lambda'}\mu}{g'\mu}$, obtaining fidelity  $F_{K}^{\rm prob}\rightarrow\dfrac{{\lambda'}+\mu+{\lambda'}\mu}{g'^{2}(\mu+1)}$ in the limit $K\rightarrow\infty$.
\end{enumerate}
Since the limit values coincide with the optimal probabilistic fidelities of Eq. (\ref{prob-opt}), we conclude that the noiseless amplifier, for suitable values of the amplification parameter, is optimal for the probabilistic amplification and purification of noisy coherent states.

\section{Evaluation of the quantum benchmark}\label{app:CFT}

The CFT for joint amplification  and purification can be upper bounded as
\begin{equation}\begin{split}\label{cft-prob-gen}
F_{g'}^{prob,c}&=\|\Gamma_{\sigma}\|_{\times}=\|\Gamma_{\sigma}^{\mathrm{T}_{2}}\|_{\times}\\&\le \|\Gamma_{\sigma}^{\mathrm{T}_{2}}\|_{\infty}\\
&=c_1\left \| \int\dfrac{d^{2}\alpha}{\pi} 
D(c_2 \alpha)x'^{a^\dag a}D^\dag(c_2 \alpha)\otimes|g'\alpha\>\<g'\alpha|\right\|_\infty\\
&=\dfrac{{\lambda'}+\mu+{\lambda'}\mu}{{\lambda'}+\mu+{\lambda'}\mu+g'^{2}(\mu+1)} \, ,
\end{split}\end{equation}
where $T_2$ is the partial transpose over the Hilbert space of the input state and $c_1$, $c_2$ and $x'$ are: 
\begin{align*}
&c_1=\frac{{\lambda'}\mu+{\lambda'}+\mu}{\mu+1}\\
&c_2=\frac{\sqrt{({\lambda'}\mu+{\lambda'}+\mu)({\lambda'}+\mu)}}{\mu}\\
&x'=\frac{{\lambda'}\mu+{\lambda'}+\mu}{({\lambda'}+\mu)(\mu+1)}  \, .
\end{align*}

Here we show that the upper bound can be achieved by the  deterministic measure-and-prepare channel
\begin{equation}
\widetilde{C}(\rho)=\int\dfrac{d^{2}\beta}{\pi}\langle\beta|\rho|\beta\rangle |z  \beta\rangle\langle  z \beta|  \qquad z=\frac{g'\mu}{{\lambda'}+\mu+{\lambda'}\mu} \, .
\end{equation}

Explicitly,  the average fidelity is 
\begin{align}
F^{{\rm M\&P}}_{g'}&\nonumber=\int\frac {d^2\alpha}{\pi}\dfrac{d^{2}\beta}{\pi}{\lambda'} e^{-{\lambda'}|\alpha|^2}\langle\beta|\rho_{\mu,\alpha}|\beta\rangle|\<g'\alpha |z  \beta\rangle|^2\\
&=\nonumber \dfrac{{\lambda'}\mu}{({\lambda'}\mu+{\lambda'}+\mu)g'^{2}z^{2}-2\mu g'^{2}z+\mu({\lambda'}+g'^{2})} \\
&=\dfrac{{\lambda'}+\mu+{\lambda'}\mu}{{\lambda'}+\mu+{\lambda'}\mu+g'^{2}(\mu+1)} \, .
\end{align}
Note that this value coincides with the upper bound (\ref{cft-prob-gen}).   Hence, we conclude that the CFT amplification is
\begin{equation}
F_{g'}^{\rm det,c} = F_{g'}^{prob,c}=\dfrac{{\lambda'}+\mu+{\lambda'}\mu}{{\lambda'}+\mu+{\lambda'}\mu+g'^{2}(\mu+1)}\, .
\end{equation}
Eq.(\ref{cft explic n}) follows by expressing the above equation in terms of the expected photon numbers $N_{\rm C}$ and $\widetilde N_{\rm T}$.

\end{document}